\documentclass[11pt]{article}
\usepackage[british]{babel}
\usepackage[margin=1in]{geometry}
\usepackage{setspace}
\usepackage{graphicx}
\usepackage{booktabs}
\usepackage{longtable}
\usepackage{array}
\usepackage{float}
\usepackage{amsmath,amssymb}
\usepackage{caption}
\usepackage{subcaption}
\usepackage{multirow}
\usepackage{lscape}
\usepackage{threeparttable}
\usepackage{adjustbox}
\usepackage{tabularx,makecell}
\usepackage{tikz}
\usepackage{siunitx}
\usepackage[hidelinks]{hyperref}
\usepackage[backend=biber,style=apa,natbib=true]{biblatex}
\DeclareLanguageMapping{british}{british-apa}
\newcommand{\WDTitle}{Droughts and Deluges: Non-Linear Effects of Climate Extremes on the Gender Gap in Labour Supply}

\newcommand{\sym}[1]{\ifmmode^{#1}\else\(^{#1}\)\fi}

\begin{document}
{\Large\bfseries \WDTitle\par}
\vspace{0.75cm}

\begin{center}
    Jheelum Sarkar\footnote{PhD Candidate, Department of Economics, American University. Email: js8622a@american.edu}
\end{center}

\begin{abstract}
\noindent Over the past three decades, extreme climate events have caused losses of worth USD 4.5 trillion. Using collective bargaining model, I find that the gendered labour supply response to adverse shocks is not straightforward since it depends on relative strength of income and substitution effects of men's and women's participation. Using a panel of 151 countries (1995-2019), I examine how extreme climate conditions shape gender gap in labour force participation. This study finds that the gender gap in paid labour exhibits a U-shaped relationship with droughts and an inverted U-shaped relationship with extreme wet conditions. The drought pattern is primarily driven by gender gap in employment while wetness affects gender gap in participation through unemployment. These relationships vary with country characteristics. Countries with high disaster-displacement risk exhibit declining gender gaps in participation during excess wetness while moderate-risk economies experience expanded gaps during droughts. Furthermore, the drought U-shape is most pronounced in countries with low to moderate empowerment while the nonlinear wet responses is concentrated only in moderately empowered countries. Lastly, both droughts and excess wetness expands gender gap in countries with weak net resilience to climate shocks.
\end{abstract}

\section{Introduction}
According to World Meteorological Organisation (\citeyear{WMO_GADCU_2024_2028_PreventionWeb}), global average temperature is likely to exceed the 1.5°C threshold in the next five years. This implies that extreme climate events such as floods, droughts and storms are more frequent and intense (\cite{WMOExtremeWeather}). By 2025, droughts have adversely affected 1.5 billion lives worldwide (\cite{UNDRR_GAR2025_Droughts}). Floods account for nearly 50\% of all extreme weather events over four decades (\cite{UNDRR_GAR2025_Floods}). Eastern and Southern Africa are major hotspots for prolonged droughts where more than 90 million people suffer from drought-induced acute hunger (\cite{NDMC_UNCCD_DroughtHotspots_News_2025}). 23\% of the global population are exposed to floods and majority of them are located in East and South Asian countries (\cite{Rentschler2022}). Even advanced economies like U.S. and U.K. are exposed to floods and droughts (\cite{WahlZoritaDiazHoell2022SouthwesternUSDrought,ChuEtAl2025FloodsMortalityTriplyRobust}).

Extreme weather shocks affect labour supply by altering productivity, sectoral composition, migration and time reallocation (\cite{ParsonsEtAl2021LaborLossesAdaptationPotential, ChenYang2019TemperatureOutputChina, GroenPolivka2008Katrina, ZivinNeidell2014TimeAllocation}). In India, for instance, extreme temperature increased absenteeism in manufacturing sector (\cite{SomanathanEtAl2021}). During 1988-2019, flood risks reduced labour supply by 0.33\% in United States (\cite{JiaMaXie2022_NBER30250}). However, evidence is mixed when it comes to gender disparities in paid labour. Droughts reduced women's agricultural working days by 19\% relative to men's in rural India (\cite{AfridiMahajanSangwan2022GenderedEffectsDroughts}). On the contrary, both negative and positive rainfall shocks have no significant effect on gender gap in paid labour in India (\cite{MaitraTagat2024LaborSupplyRainfallShocks}). After a flood, women spent more time in paid labour than men in Bangladesh (\cite{VitellozziGiannelli2024ThrivingInTheRain}). Despite rising trend in women's participation after flood in Pakistan, it remains much lower than men (\cite{Akter2021CatastrophicFloodsGenderDivisionLaborPakistan}).  

These mixed findings could be attributable to a non-linear relationship between extreme events and gender differences in paid labour. Mild droughts can incentivize both men and women to increase paid work for coping with financial hardships while acute droughts may raise demand for home production (\cite{MaitraTagat2024LaborSupplyRainfallShocks,AfridiMahajanSangwan2022GenderedEffectsDroughts,ElmallakhEtAl2025PRWP11191}). Moderate wet conditions increases demand for unpaid care work due to disease burdens and can expand gender inequality in paid labour (\cite{DimitrovaEtAl2023PNAS,Boyd2023RainfallTimeUseUganda, DimitrovaEtAl2022LancetPlanetHealthPrecipVar}). But extreme wet conditions trigger floods and storms which could induce both men and women to temporarily work more for smoothing household consumption and distress-driven labour supply (\cite{GrogerZylberberg2016AEJApplied,PechaGarzon2017IDBInformality,VitellozziGiannelli2024ThrivingInTheRain}). Using a panel of 151 countries over 1995-2019, this paper, therefore, examines whether drought and extreme wet conditions share non-linear relationship with gender gap in labour force participation (LFP). To focus on extreme climate conditions, I construct extreme dry and wet intensities using 12-month Standardized Precipitation Evapotranspiration Index (SPEI).  

Using a two way fixed effects (TWFE) model, three interesting results emerge. Firstly, gender gap in LFP shares a non-linear relationship with extreme climate conditions: drought intensity has a U-shaped relation with gender gap in LFP while wet intensity exhibits inverted U-shaped relationship. Secondly, excess wetness mainly affects gender gap in unemployment while drought conditions shifts gender gap in LFP through employment. Lastly, the key patterns differ by country characteristics, namely, natural disaster-related displacement risk, women's empowerment and net resilience. Countries with high displacement risk are more sensitive to excess wetness while drought responses are pronounced in countries with moderate displacement risk. Droughts and excess wet conditions influence gender gap in LFP in countries with low and moderate empowerment. Least resilient countries exhibit U-shaped relationship between droughts and gender gap in paid labour. Main findings remain robust under alternative specifications.

Rest of the paper is organized as follows. I set up a theoretical framework in section~\ref{theory}. Section~\ref{data} describes the data and  estimation strategy. Section~\ref{results} presents the main results, robustness checks and heterogeneous effects. This is followed by concluding remarks in section~\ref{conclusion}.

\section{Theoretical Framework}\label{theory}
I develop a simple theoretical model to discuss why the effect of climate extremes on the gendered paid labour is not straightforward. Because labour supply decisions and time allocation often depend on household wealth, spouse's education and income, I am setting up a collective bargaining household model. Apart from changing household resources, a climate shock alters the relative returns and costs of market work. This is reflected in the income and substitution effects. The net balance between these two effects determine the direction of the gendered labour supply response to a climate shock.

\paragraph{Setup.}
Let us consider a representative household with husband ($m$) and wife ($f$). Each spends one unit of time across paid work ($L_j$), home production ($h_j$), and leisure ($l_j$), for $j\in\{f,m\}$. That is,
\begin{equation}\label{eq:time}
    L_i+\ell_i+h_i=1
\end{equation}
Let $U_m$ and $U_f$ denote husband's and wife's utility functions respectively which depend on consumption ($C_i$), leisure ($\ell_i)$ and home production ($H$):
\begin{equation}
U_m=U_m(C_m, \ell_m, H), 
\end{equation}
\begin{equation}
    U_f=U_f(C_f, \ell_f, H)-\kappa(L_f;\theta)
\end{equation}
where  $\quad U_i'(\cdot)>0,\;U_i''(\cdot)<0$. $\kappa(.)$ is social cost of the woman's paid labour due to social norms (e.g., mobility, domestic work and care burdens). $\kappa$ is also sensitive with climate shock, given by parameter $\theta$. Higher the value of $\theta$, more adverse is the shock. I parameterize $\kappa(L_f;\theta)$ as: 
\begin{equation}\label{eq:kappa_param}
\kappa(L_f;\theta)= s(\theta)\,\tilde{\kappa}L_f,
\qquad s(\theta)>0,\quad s'(\theta)\ge 0,\quad \kappa'(L_f)>0\,\quad \kappa''(L_f)\ge 0
\end{equation}

Consumptions ($C_f, C_m$) are financed by the labour income and non-labour earnings from asset ownership. Thus, the usual household budget constraint is:
\begin{equation}\label{eq:budget}
    C_f+C_m=w_fL_f+w_mL_m+Y
\end{equation}
\noindent where $w_j$ denotes the wages and $Y$ captures non-labour income (e.g., farm profits, transfers, asset income). In presence of negative shock such as floods and droughts, physical assets ($Y'(\theta)<0$) are destroyed and wages can fall ($w_i'(\theta)<0$).

Home production function produces good $H$ (e.g., domestic and care work) which depends on contributions by both family members and shock parameter. An extreme climate event ($\theta$) can increase the amount of time required to complete household responsibilities (e.g., water collection, caregiving), or can reduce the productivity of time spent on household work. It is given by 
\begin{equation}\label{eq:Hprod}
H=H(h_m,h_f;\theta), \qquad H_{h_j}>0,\;\; H_{h_jh_j}<0.
\end{equation}

Using collective bargaining model approach, the utility maximization of the representative household is given by:
\begin{equation}
    U=\alpha U_f+(1-\alpha)U_m 
\end{equation}
 subject to equations~(\ref{eq:time}), (\ref{eq:budget}) and (\ref{eq:Hprod}). Solving this maximization problem gives:
\paragraph{Opportunity costs of home production.} 
Let $\lambda$ denote the Lagrange multiplier. Define the shadow value of an hour spent in home production by spouse $j$ as
\begin{equation}\label{eq:shadow_home}
q_j(\theta)\;\equiv\;\frac{u_H}{\lambda}\,H_{h_j}(h_m,h_f;\theta),\qquad j\in\{f,m\}.
\end{equation}
\noindent where $u_H$ is marginal utility from household production. 
From the F.O.Cs,
\begin{equation}\label{eq:opp_home_f}
\begin{aligned}
u_HH_{h_f} &= \lambda w_f-\alpha\kappa,\\
\implies\quad q_f(\theta) &= w_f-\frac{\alpha\kappa}{\lambda}.
\end{aligned}
\end{equation}

and
\begin{equation}\label{eq:opp_home_m}
q_m(\theta)=w_m.
\end{equation}
Thus, women's value of household work her wage net of the social cost from working outside. On the contrary, men's opportunity cost of home labour is simply their wages.\paragraph{Reduced-form labour supply and decomposition.}Solving the household problem yields Marshallian labour-supply functions:
\begin{align}
\frac{dL_f^*}{d\theta}
&=
\underbrace{\frac{\partial \mathcal{L}_f}{\partial Y}\,Y'(\theta)}_{\text{Income effect}}
\;+\;
\underbrace{\left(
\frac{\partial \mathcal{L}_f}{\partial w_f}\,w_f'(\theta)
+\frac{\partial \mathcal{L}_f}{\partial w_m}\,w_m'(\theta)
\right)}_{\text{Substitution effect}}
\;+\;
\underbrace{\frac{\partial \mathcal{L}_f}{\partial s}\,s'(\theta)}_{\text{Social cost}}
\;+\;
\underbrace{\frac{\partial \mathcal{L}_f}{\partial\theta}}_{\text{Home prod.}},
\label{eq:income_sub_f}\\[16pt]
\frac{dL_m^*}{d\theta}
&=
\underbrace{\frac{\partial \mathcal{L}_m}{\partial Y}\,Y'(\theta)}_{\text{Income effect}}
\;+\;
\underbrace{\left(
\frac{\partial \mathcal{L}_m}{\partial w_f}\,w_f'(\theta)
+\frac{\partial \mathcal{L}_m}{\partial w_m}\,w_m'(\theta)
\right)}_{\text{Substitution effect}}
\;+\;
\underbrace{\frac{\partial \mathcal{L}_m}{\partial\theta}}_{\text{Home prod.}}.
\label{eq:income_sub_m}
\end{align}

In each of the above equation, 
%\vspace{-0.5em}
\begin{itemize}
    \item \emph{Income Effect:} This is captured by the non-labour income channel. Higher household wealth reduces women's labour supply, i.e., $\frac{\partial \mathcal{L}_f}{\partial Y}<0$ (\cite{Fu_the_2016, AlAzzawi_household_2019, Klasen_what_2015}). While there isn't adequate research on how men's labour supply is linked to household wealth, Cesarini et al. (\citeyear{Cesarini_the_2015}) shows that higher non-labour income reduces labour of both men and women, i.e., $\frac{\partial \mathcal{L}_m}{\partial Y}<0$. If adverse shock depletes non-labour wealth and physical assets, $Y'(\theta)<0$. Thus, the income effect is positive for both men and women.
    \[
    \frac{\partial \mathcal{L}_i}{\partial Y}\,Y'(\theta) \;>\; 0
    \]
    \item \emph{Substitution Effect:} This is measured by channel or changes in wages. In case of both men and women, an increase in own wage raises their own paid labour ($\frac{\partial \mathcal{L}_i}{\partial w_i}>0$). But the cross-wage effects differ by gender. For women, higher husband's income reduces their labour supply, i.e., $\frac{\partial \mathcal{L}_f}{\partial w_m}<0$ \citep{zhu2023husbands_wp, albanesi2022slowing, Cheng_relative_2023}. On the contrary, some studies suggest that higher wife's income can induce husbands to work more, i.e., $\frac{\partial \mathcal{L}_m}{\partial w_f}>0$ \citep{Cheng_relative_2023, chang2025spousal}. If negative shock may dampen wages ($w_i'(\theta)<0$), the market-return effect is negative for men but ambiguous in case of women since the latter depends on the net strength of own wage and cross-wage effects.
\item \emph{Home Production Channel:} 
This channel captures the effect of climate shocks on labour supply through the home production lens $H(h_m,h_f;\theta)$. It operates  in two opposing ways. First, adverse shocks may increase the time required for household chores (e.g., water collection, caregiving, or repairing damaged assets). This raises the marginal value of home time $H_{h_j}$ by reallocating an individual's time toward home production and away from paid labour. Second, shocks may reduce the productivity of home production (e.g., through damage to household assets or infrastructure) because it since lowers the returns to time spent in home activities. This can potentially shift an individual's time more towards market work. These effects operate through changes in the shadow value of home time and are captured by $\frac{\partial \mathcal{L}_i}{\partial \theta}$. The net impact depends on which force dominates:
\[
\frac{\partial \mathcal{L}_i}{\partial\theta} \;\gtrless\; 0.
\]
\item \emph{Social Cost Channel.} Adverse shock such as climate extremes increase social cost of women's paid labour ($s'(\theta)>0$) by raising mobility constraints, household burden and reducing economic opportunities in affected regions\citep{AfridiMahajanSangwan2022GenderedEffectsDroughts, Escalante_assessing_2022, rao2019gendered, flato2014droughts}. Conservative social constraints, as indicated by higher $s(\theta)$, raises marginal cost of women's paid work, i.e., $\frac{\partial \mathcal{L}_f}{\partial s}<0$. This generates an additional substitution for paid labour in case of women through $s(\theta)$. Hence, the climate shock reduces women's paid labour through social cost channel: 
        
        \[\frac{\partial \mathcal{L}_f}{\partial s}\,s'(\theta)<0
        \]
    \end{itemize}

The effect of climate shock on gender gap in paid labour supply is given by:
\begin{equation}
\begin{aligned}
\frac{d\,\mathrm{Gap}}{d\theta}
&= \frac{dL_m^*}{d\theta} - \frac{dL_f^*}{d\theta} \\
&=
\underbrace{A\,Y'(\theta)}_{\text{Income effect}}
+\underbrace{
\overbrace{B\,w_f'(\theta)+C\,w_m'(\theta)}^{\text{Own-wage and cross-wage effect}}
}_{\text{Substitution effect}}
-\underbrace{\frac{\partial \mathcal{L}_f}{\partial s}\,s'(\theta)}_{\text{Social cost}}
+\underbrace{D'(\theta)}_{\text{Home production effect}}.
\end{aligned}
\end{equation}
Given that social cost worsens after adverse shock, the direction in which climate shock affect gender gap in paid labour depends on the difference between men's and women's income, substitution and home production effects.
If higher non-labour income reduces women's labour supply more than that of men's paid labour (i.e.,  $\frac{\partial \mathcal{L}_f}{\partial Y}<\frac{\partial \mathcal{L}_m}{\partial Y}$), $A>0$. Because $w'_i(\theta)<0$, men's labour supply increases due to rise in their own-wage and spouse's wage (i.e., $B\geq0$). Thus, substitution effect is negative for men. On the contrary, own-wage and husband's wage have opposite effects on women's labour. Studies have shown that women reduce their participation if their husbands earn more. That is, $\frac{\partial\mathcal{L}_f}{\partial w_m}<0$. Thus, women's substitution effect depends on relative strength of own-wage and cross-wage effect ($B \gtrless 0$). Existing studies have shown that women bears more burden of household work after climate shock. This implies that $D'(\theta)>0$. Hence, the effect of climate shock on gender gap in paid labor depends on income and substitution effects.

\section{Data and Research Method}\label{data}
This study uses a panel of 151 countries (1995-2019) on paid labour, economic and climate conditions.
\vspace{-1.0\baselineskip}
\paragraph{Labour Data.} Annual gender-disaggregated data on labour force participation, employment rate and unemployment rate are drawn from ILOSTAT modeled estimates in each year. ILO-modeled estimates are useful for making cross-country comparisons since these use harmonised methods across countries over years. The modeled series include imputed values where underlying national estimates are missing. Because imputation is done only using observations that are sufficiently comparable across countries, these estimates are reliable. 
\vspace{-1.0\baselineskip}
\paragraph{SPEI data.} The Standardized Precipitation Evapotranspiration Index (SPEI) is a multi-scalar metric for drought and excess wetness. It standardizes anomalies in climate water balance that allow comparisons across regions over time. Monthly gridded SPEI values are obtained from the \texttt{SPEIbase v2.10} and aggregated to the country levels over 1980–2019. \texttt{SPEIbase v2.10} is a global dataset which is provided at (\SI{0.25} x \SI{0.25}) resolution. SPEI-12 is used in this analysis to derive drought and extreme wet intensities which are main independent variables. The rationale behind focusing on 12-month SPEI instead of 3-month and 6-month SPEI values is that it captures climatic shifts as reflected in the long-term changes in moisture contents while the shorter SPEI scales (3 or 6-month SPEI) capture seasonal and agricultural shocks respectively (\cite {VicenteSerranoEtAl2010,BegueriaEtAl2010SPEIbase}). Negative and positive SPEI-12 values indicate drier and wet conditions respectively. 

\begin{figure}[htbp]
    \centering
    \includegraphics[width=0.7\linewidth]{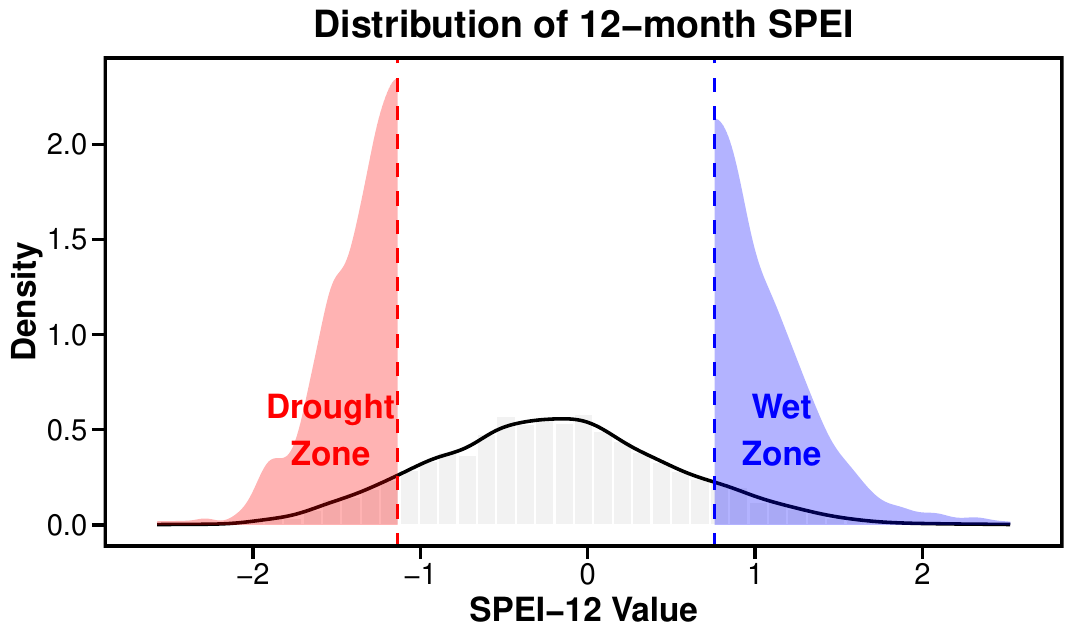}
    \caption{\textit{Distribution of 12-month SPEI during 1995-2019. 10th percentile and 90th percentile values are considered as thresholds for extreme drought and wet intensities.}}
    \label{fig:1}
\end{figure}

Let $SPEI_{ct}$ denote SPEI-12 value for country $c$ and year $t$. Figure~\ref{fig:1} shows the distribution of $SPEI_{ct}$.  

To capture extreme drought and wet conditions, I define:\\
(i) \textbf{Drought intensity:} the extent to which the distance falls behind the 10th percentile and zero otherwise.
\begin{equation}
\mathrm{Drought}_{ct}=\max\{0,\,-SPEI_{ct}-t_D\},
\end{equation}
\noindent where $t_D$ is the 10th percentile value in the sample $SPEI_{ct}$ distribution; $\mathrm{max(,)}$ function captures extreme droughts since $\mathrm{Drought}_{ct}$ takes positive values only when $SPEI_{ct}<t_D$.\\
(ii) \textbf{Wet intensity:} the extent to which actual magnitude of 12-month SPEI exceeds its 90th percentile value and zero otherwise.
\begin{equation}
\mathrm{Wet}_{ct}=\max\{0,\,SPEI_{ct}-t_W\},
\end{equation}
\noindent where $t_W$ is the 90th percentile value in the sample $SPEI_{ct}$ distribution; $\mathrm{max(,)}$ function captures extreme wet conditions since $\mathrm{Wet}_{ct}$ takes positive values only when $SPEI_{ct}>t_W$. 

\paragraph{Demographic and Economic Indicators.} Data on demographic and economic controls are drawn from World Bank's World Development Indicators. These include GDP growth, foreign direct investment (FDI) inflows, gender-disaggregated population, land area, fertility, share of dependent population and urbanization. Using these data, I construct sex ratio and population density for each country in every year. Economic growth, urbanization and FDI also influence labour force participation but the same year values are likely to cause reverse causality. Hence, I use lagged values of economic growth, urbanization and FDI inflows as controls.

Data on disaster displacement risk is obtained from the Internal Displacement Monitoring Centre. As the term suggests, it measures vulnerability to displacement due to natural disasters. Women's empowerment data comes from the World Bank’s Women, Business and the Law database. It measures legal and institutional constraints that shape women’s agency and access to work. I measure a country's net resilience by the Notre Dame Global Adaptation Initiative (ND-GAIN) Country Index. It highlights net capacity to withstand and adapt to climate shocks.

Table~\ref{tab:summstats} shows the summary statistics. 
\begin{table}[]\centering
\caption{Summary Statistics}
\label{tab:summstats}
\begin{threeparttable}
\small
\begin{tabular}{lccc}
\toprule
Variable & $N$ & Mean & Std. dev. \\
\midrule
Gender gap in LFP (pp)                & 4,000 & 21.620 & 14.424 \\
Gender gap in employment (pp)         & 4,000 & 20.691 & 13.796 \\
Gender gap in unemployment (pp)       & 4,000 & -1.847 & 3.558  \\
Drought intensity (SPEI-12 tail)      & 5,400 & 0.021  & 0.098  \\
Wet intensity (SPEI-12 tail)          & 5,400 & 0.024  & 0.120  \\
Economic growth (lag)                 & 4,865 & 3.776  & 6.080  \\
Urbanization (lag)                    & 5,160 & 57.339 & 24.529 \\
FDI inflows (lag, \% of GDP)          & 4,596 & 0.064  & 0.453  \\
Age dependency ratio                  & 5,375 & 62.221 & 18.981 \\
Fertility rate                        & 5,375 & 2.989  & 1.591  \\
Sex ratio (female/male)               & 5,375 & 1.012  & 0.083  \\
Population density                    & 5,236 & 400.541 & 1869.108 \\
\bottomrule
\end{tabular}
\begin{tablenotes}\footnotesize
\item Notes: Table reports means and standard deviations for the main variables used in the analysis. Gender gap in labour outcomes is the difference between men's and women's outcomes. Gender gaps are expressed in percentage points. Lagged variables are denoted by $t\!-\!1$.
\end{tablenotes}
\end{threeparttable}
\end{table}
\subsection{Empirical Methodology}
Using a TWFE model, I explore the relationship between climate extremes and gender gap in paid labour using the following specification:
\begin{equation}\label{main_equation}
    \mathrm{GenderGap}_{ct}= \beta_0+\beta_1\mathrm{Drought}_{ct}+\beta_2\mathrm{Drought}^2_{ct}+\beta_3\mathrm{Wet}_{ct}+\beta_4\mathrm{Wet}^2_{ct}+\mathbf{X'}_{ct}\beta_5+\mathbf{Z'}_{ct-1}\beta_6+\tau_c+\tau_t+u_{ct},
\end{equation}
\noindent where $\mathrm{GenderGap}_{ct}$ is the difference between men's and women's LFP. $\mathrm{Drought}_{ct}$ and $\mathrm{Wet}_{ct}$ are respectively drought and wet intensities. If the coefficients of squared terms $\mathrm{Drought}^2_{ct}$ and $\mathrm{Wet}^2_{ct}$ are showing opposite sign to their linear counterparts and are statistically significant, that captures the non-linear relationship between climate extremes and gender gap in paid labour outcomes. $\mathbf{X'}_{ct}$ are vectors of time-varying controls which include fertility, share of dependent population and gender gap in secondary education enrollment in country $c$ and year $t$. $\mathbf{Z'}_{ct-1}$ are vectors of lagged controls including economic growth, urbanization and FDI. $\tau_c$ and $\tau_t$ are country and year fixed effects. $u_{ct}$ are standard errors which are clustered at the country level.\\
\section{Results}\label{results}
\subsection{Main Results}
\begin{figure}
    \centering
    \includegraphics[width=0.8\linewidth]{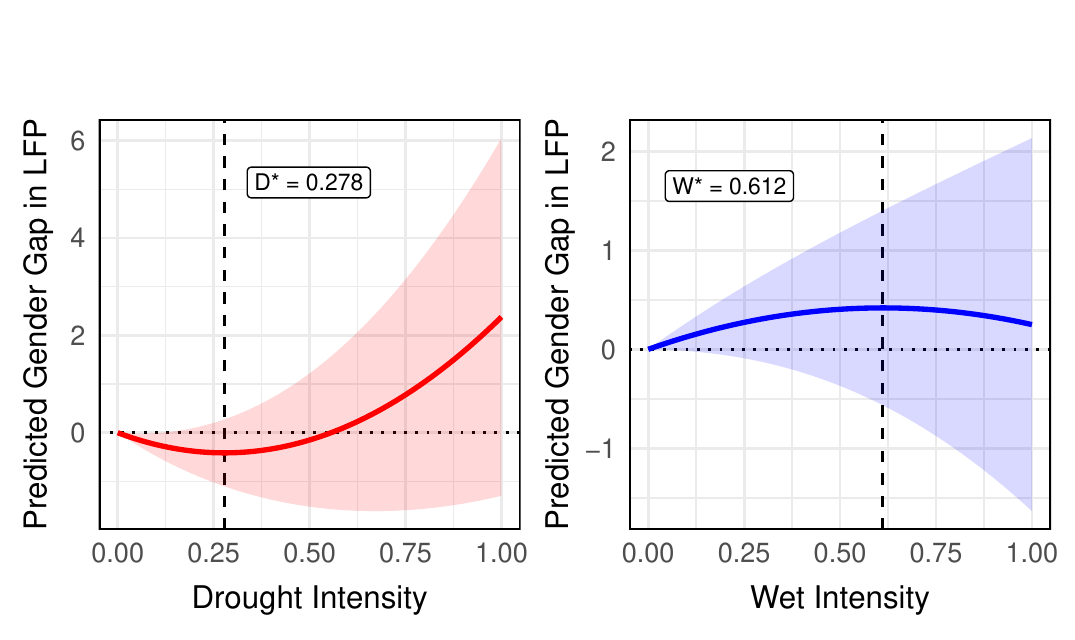}
    \caption{\textit{Non-linear Relationship Between Climate Extremes and Gender Gap in LFP. Plots are based on the predicted gender gap in LFP using eq~\ref{main_equation}. Left panel captures the relationship with extreme drought and right panel shows the relationship with extreme wet conditions. Dashed vertical lines mark the estimated turning points: $D^*=0.278$ and $W^*=0.612$. Shaded areas indicate 95\% confidence intervals and the horizontal dotted line denotes zero.}}
    \label{fig:3}
\end{figure}

Table~\ref{tab:climate_gendergaps_checks} presents the main estimates from equation~(\ref{main_equation}). Column (1) reports results for the main dependent variable, gender gap in labour force participation (LFP). Based on these estimates, Figure~\ref{fig:3} visualizes the fitted relationship between gender gap in LFP with drought intensity (left panel) and wet intensity (right panel). Columns (2) and (3) report estimates for gender gap of LFP's constituents, namely, employment rate and unemployment rate.  

\textbf{Droughts.} Column~(1) shows non-linear relationship between gender gap in LFP and drought intensity (since $\beta_1=-2.977$ and $\beta_2=5.354$). As drought intensity increases from zero to its turning point ($D^{*}=0.278$), gender gap in LFP shrinks by nearly 3 percentage points (Column (1), Table~\ref{tab:climate_gendergaps_checks}). Beyond this threshold, the gap expands (Column (1), Table~\ref{tab:climate_gendergaps_checks}; Figure~\ref{fig:3}). Thus, gender gap in LFP has a U-shaped relationship with drought intensity (Figure~\ref{fig:3}). Drought conditions primarily affect gender gap in LFP through increasing gender gap in employment (Column (2), Table~\ref{tab:climate_gendergaps_checks}). This pattern explains mixed findings in the literature: moderate dry spells increases women's paid labour while large scale drought tend to the widen gender gap (\cite{AfridiMahajanSangwan2022GenderedEffectsDroughts, MaitraTagat2024LaborSupplyRainfallShocks}).

\textbf{Excess Wetness.} Wet intensity also depicts non-linear effects. At low wetness, gender gap initially expands by 1.37 percentage points but it is not statistically significant (Column (1), Table~\ref{tab:climate_gendergaps_checks}). As wet intensity exceeds $W^{*}=0.612$, gender gap starts to decline (Column (1), Table~\ref{tab:climate_gendergaps_checks}; Figure~\ref{fig:3}). This implies that there exists inverted U-shaped relationship between gender gap in LFP and wet intensity (Figure~\ref{fig:3}). It is noteworthy that the post-threshold fall in gender gap is attributable to declining gender gap in unemployment
(Column (3), Table~\ref{tab:climate_gendergaps_checks}). This pattern is consistent with recent findings that women increases their paid labour supply after floods (\cite{VitellozziGiannelli2024ThrivingInTheRain, Akter2021CatastrophicFloodsGenderDivisionLaborPakistan}). 
\vspace{-0.6\baselineskip}

\subsection{Robustness Checks}
\vspace{-0.5\baselineskip}
One concern with the baseline specification is that the non-linear relationship between climate extremes and gender gap in paid labour could be attributable to quadratic form of droughts and excess wetness in equation~(\ref{main_equation}). Imposing squared terms may generate non-linear curvature and mask the true nature of the relationship. Hence, I introduce two alternative specifications by relaxing quadratic assumptions while keeping everything else unchanged in equation~(\ref{main_equation}). 

\begin{figure}[htbp]
    \centering
    \includegraphics[width=0.8\linewidth]{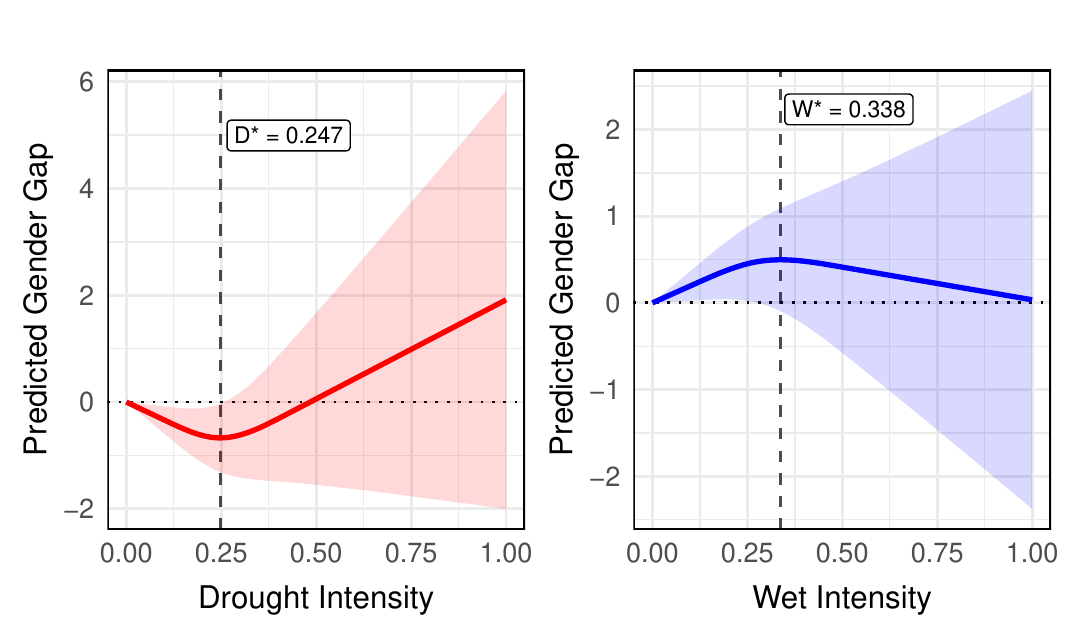}
    \caption{\textit{Non-linear Relationship Between Climate Extremes and Gender Gap in LFP. Plots are based on the predicted gender gap in LFP using cubic spline specification. Left panel captures the relationship with extreme drought and right panel shows the relationship with extreme wet conditions. Dashed vertical lines mark the estimated turning points: $D^*=0.247$ and $W^*=0.338$. Shaded areas indicate 95\% confidence intervals and the horizontal dotted line denotes zero.}}
    \label{fig:spline}
\end{figure}

\paragraph{Cubic Spline Specification.} 
\vspace{-0.5\baselineskip}
First, I model drought intensities and wetness intensities using spline bases instead of quadratic form of climate extreme measures in equation~(\ref{main_equation}). This approach model non-linear trends by altering both the number and position of knots which join different polynomial segments. By doing this, splines provide more continuous and accurate approximations to the true functional form. Using the distribution of drought and wet intensities separately, I use 25th, 50th and 75th percentiles as three interior knots. This generates two spline terms which are now main independent variables: $\mathrm{DroughtSpline}_{ct,1}$ and $\mathrm{DroughtSpline}_{ct,2}$ for drought, and $\mathrm{WetSpline}_{ct,1}$ and $\mathrm{WetSpline}_{ct,2}$ for wetness. Results are reported in table~\ref{tab:robust_spline_3cols} and I summarize the resultant relationship using predicted response plots in figure~\ref{fig:spline}. These yield similar pattern as the main results which implies that the non-linear relationship in figure~\ref{fig:3} is not merely because of quadratic specification.

\paragraph{Piecewise Linear Specification.}
\vspace{-0.5\baselineskip}
Next, I replace the quadratic forms of droughts and excess wetness in equation~(\ref{main_equation}) with piecewise linear specification. The objective is to test whether the non-linearity exists, that is, the slope changes sign around the baseline turning points. Using $D^{*}=0.278$ and $W^{*}=0.612$, I construct $D^{\text{below}}_{ct}=\min\{\mathrm{Drought}_{ct},D^{*}\}$ and $D^{\text{above}}_{ct}=\max\{0,\mathrm{Drought}_{ct}-D^{*}\}$, and analogously for wetness.

\begin{figure}[htbp]
    \centering
    \includegraphics[width=0.8\linewidth]{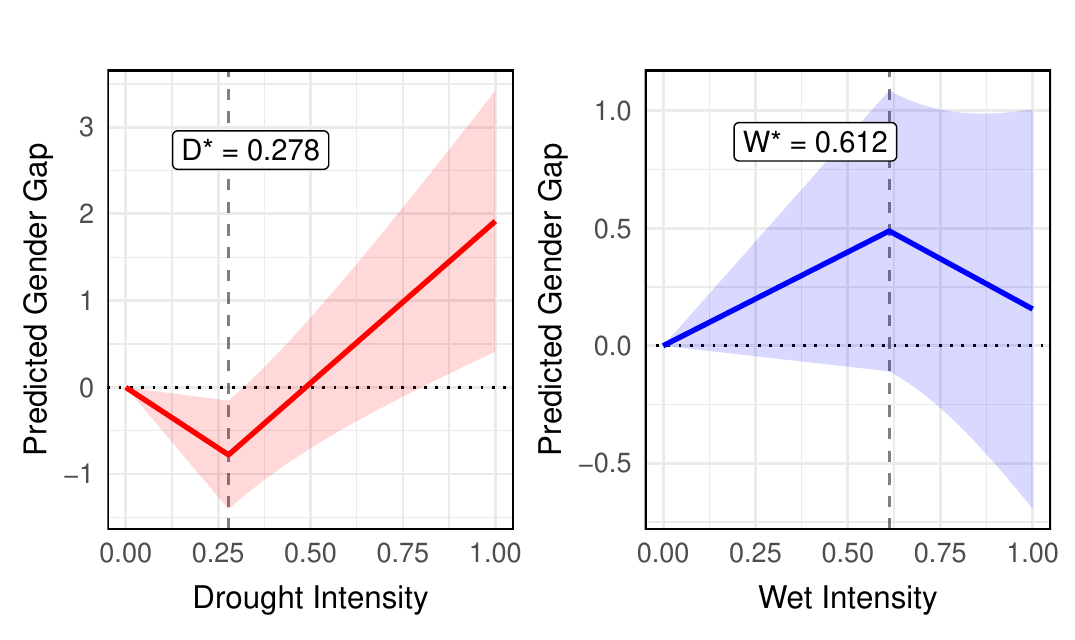}
    \caption{\textit{Non-linear Relationship Between Climate Extremes and Gender Gap in LFP. Plots are based on the predicted gender gap in LFP  using linear piecewise specification. Left panel captures the relationship with extreme drought and right panel shows the relationship with extreme wet conditions. Dashed vertical lines mark the estimated turning points: $D^*=0.278$ and $W^*=0.612$. Shaded areas indicate 95\% confidence intervals and the horizontal dotted line denotes zero.}}
    \label{fig:piecewise}
\end{figure}

The results are reported in table~\ref{tab:robust_piecewise_3cols} and figure~\ref{fig:piecewise} which remain consistent with the trends observed in figure~\ref{fig:3}. In case of droughts, gender gap initially declines below the turning point while it goes up if drought exceeds the turning point (Column~(1), Table~\ref{tab:robust_piecewise_3cols} and Figure~\ref{fig:piecewise}). This drought-led U-shaped relationship is explained by gender gap in employment which is similar to baseline results (Column~(2), Table~\ref{tab:climate_gendergaps_checks} and Column~(2), Table~\ref{tab:robust_piecewise_3cols}). In case of excess wetness, gender gap first rises if wetness falls short of the turning point but it declines once wetness goes beyond the cutoff (Column~(1), Table~\ref{tab:robust_piecewise_3cols} and Figure~\ref{fig:piecewise}). while this wetness-gender gap nexus is not statistically significant, the wetness-driven gender gap in paid labour is explained by gender gap in unemployment which remains significant as in the baseline results (Column~(3), Table~\ref{tab:climate_gendergaps_checks} and Column~(3), Table~\ref{tab:robust_piecewise_3cols}).

\subsection {Heterogeneous Effects}
\begin{figure}[htbp]
    \centering
    \includegraphics[width=0.9\linewidth]{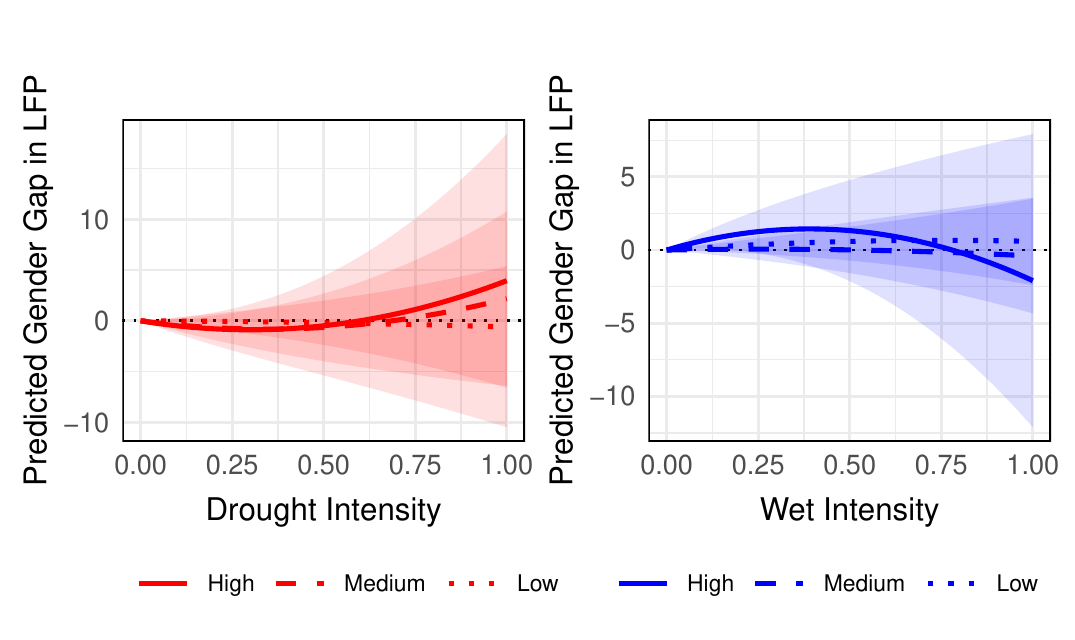}
    \caption{\textit{Heterogeneous Effects of Drought and Excess Wet Intensities on Gender Gap in LFP by  Disaster Displacement Risk. Predicted gender gap in LFP are plotted for different drought intensity (left-side panel) and excess wet intensity (right-side panel). Shaded areas indicate 95\% Confidence Intervals.}}
    \label{fig:het1}
\end{figure}

The above results indicate that gender gap in paid labour are sensitive to droughts and excess wetness. What factors explain this in non-linear relationship between climate extremes and gender gap in paid labour? I analyse equation~\ref{main_equation} to examine whether the non-linear effects differ by country characteristics, namely, disaster related displacement risk, women's empowerment and net resilience. Tables~\ref{tab:climate_gendergap_disphet}--\ref{tab:climate_gendergap_ndgain_gain} report the corresponding coefficients and figure~\ref{fig:het1} gives the visualization. 

\paragraph{Disaster-related displacement risk.}
Disaster displacement risk refers to the probability that displacement at a
certain scale will occur during a specific period of time due to the
onset of a hazardous event (\cite{unisdr2018words}). 
In countries with high displacement risk, the non-linear relationship is most pronounced for excess wetness (Column~(3), Table~\ref{tab:climate_gendergap_disphet}; Panel A, Figure~\ref{fig:het1}). This implies that excess wetness is significant in countries where wet shocks are most likely to trigger displacements. With sudden onset of wet shocks, gender gap can increase due to rise in household work, childcare burdens and safety constraints (\cite{kreutzer2023disasters, fruttero2023gendered}). But as wetness become severe, men may also lose employment and the gender gap shrinks (\cite{chowdhury2022flood,VitellozziGiannelli2024ThrivingInTheRain, JiaMaXie2022_NBER30250}). Table~\ref{tab:climate_gendergaps_checks} shows that the wet-labour nexus is driven by drop in gender gap of unemployment. 

On the other hand, the effect of droughts is prominent in countries with moderate displacement risk (Column~(2), Table~\ref{tab:climate_gendergap_disphet}; Panel A, Figure~\ref{fig:het1}). This is expected because droughts are slow-onset disasters. Women are over-represented in agricultural sectors, especially in developing and less developed countries where droughts are more frequent (\cite{chatterjee2021why}). Studies have shown that migration is gendered and women are less likely to migrate than men after droughts (\cite{sultana2024climate, AfridiMahajanSangwan2022GenderedEffectsDroughts}). However, both linear and non-linear effects of drought and excess wetness are muted in countries with low-displacement risk (Column~(1), Table~\ref{tab:climate_gendergap_disphet}; Panel A, Figure~\ref{fig:het1}).

\paragraph{Women's empowerment.}
Women's empowerment is captured by Women, Business and Law (WBL) index which measures how laws and regulations affect women's economic opportunity. This index reflects both constraints and protections which influence women's economic agency such as right to work, asset ownership, free mobility. By measuring  institutional environment rather than labour market outcomes, WBL rules out concerns about reverse causality that could arise if better female labour market outcomes improve legal protections for women. %can you write this sentence better.
Higher WBL value indicates more gender equal legal rights and fewer constraints to women's work. Using WBL distribution, I classify countries into low, medium and high empowerment terciles. 

\begin{figure}[htbp]
    \centering
    \includegraphics[width=0.9\linewidth]{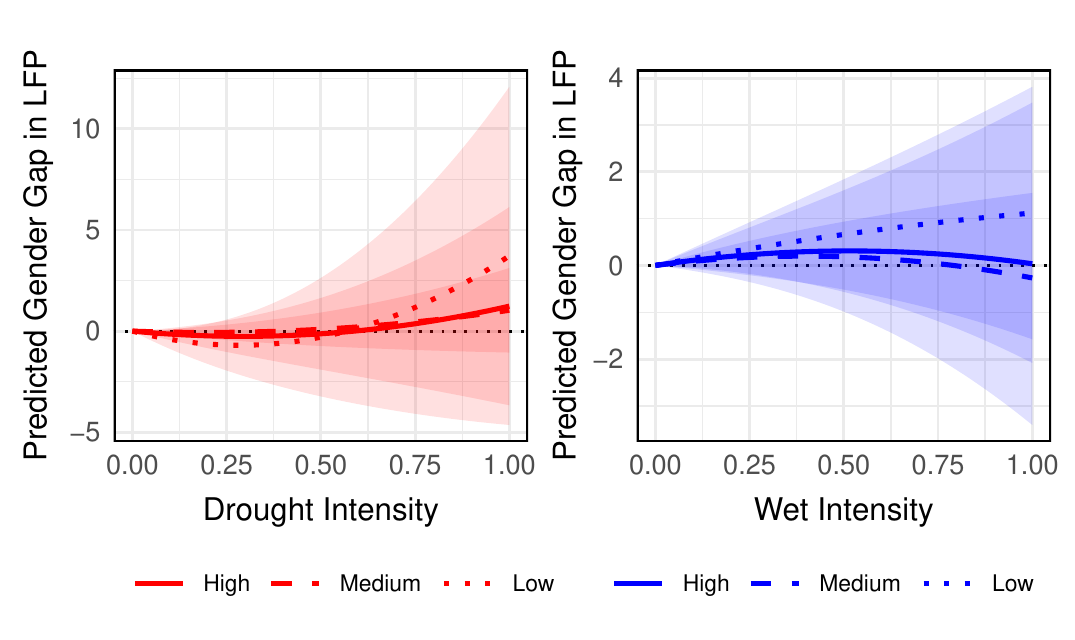}
    \caption{\textit{Heterogeneous Effects of Drought and Excess Wet Intensities on Gender Gap in LFP by Women Empowerment. Predicted gender gap in LFP are plotted for different drought intensity (left-side panel) and excess wet intensity (right-side panel). Shaded areas indicate 95\% Confidence Intervals.}}
    \label{fig:het2}
\end{figure}

The U-shaped drought-gender-gap relationship is concentrated in countries with low and moderate levels of women's empowerment (Columns~(1)–(2), Table~\ref{tab:climate_gendergap_empowerhet}; Figure~\ref{fig:het2}). This indicates that extreme weather conditions amplify gender gap in paid work where women face more legal and institutional constraints. Low drought intensities absorb more women in paid labour when households face financial loss due to dry spells (e.g., \cite{Moshoeshoe2025}). But as droughts become severe, gender norms dominate in countries with low and medium empowerment. During severe droughts, time required for unpaid domestic (e.g., water collection, fuel wood) and care work (caring for malnourished children and/or diseased elderly) increases. This, in turn, pulls women out of paid labour to cater household responsibilities (\cite{ AfridiMahajanSangwan2022GenderedEffectsDroughts, Musungu2024})

In case of excess wet conditions, the inverted U-shaped relationship is pronounced in countries with moderate empowerment (Columns~(2), Table~\ref{tab:climate_gendergap_empowerhet}; Figure~\ref{fig:het2}). At low levels of excess wetness, gender gap rises as women face increased mobility constraints and care burdens (\cite{kreutzer2023disasters}). As excess wetness goes beyond the turning point, widespread economic disruption reduces male employment and induces women to spend more time in income-generating activities \citep{VitellozziGiannelli2024ThrivingInTheRain, Akter2021CatastrophicFloodsGenderDivisionLaborPakistan,CanessaGiannelli2021}. As a result, gender inequality in paid labour declines after extreme wetness. However, gender gap is more resilient to climate extremes in countries with high empowerment (Column~(3), Table~\ref{tab:climate_gendergap_empowerhet}). 

\paragraph{Net resilience.}
I measure each country's net resilience using ND-GAIN index which is composed of vulnerability and readiness \citep{ndgain2026methodology}\footnote{$\mathrm{GAIN = (Readiness_{0\leq\mathrm{Readiness}\leq100}-Vulnerability_{0\leq\mathrm{Vulnerability}\leq1}+1)\times50}$; $0\leq\mathrm{GAIN}\leq100$}. Vulnerability is defined as a country's sensitivity, exposure and adaptive capacity to adverse effects of climate change. Readiness evaluates a country's capacity to capitalize investments for adaptation actions. It includes three components, namely, governance, economic and social readiness. Using terciles, countries are classified into low, medium and high net resilient groups. 

\begin{figure}[htbp]
    \centering
    \includegraphics[width=0.9\linewidth]{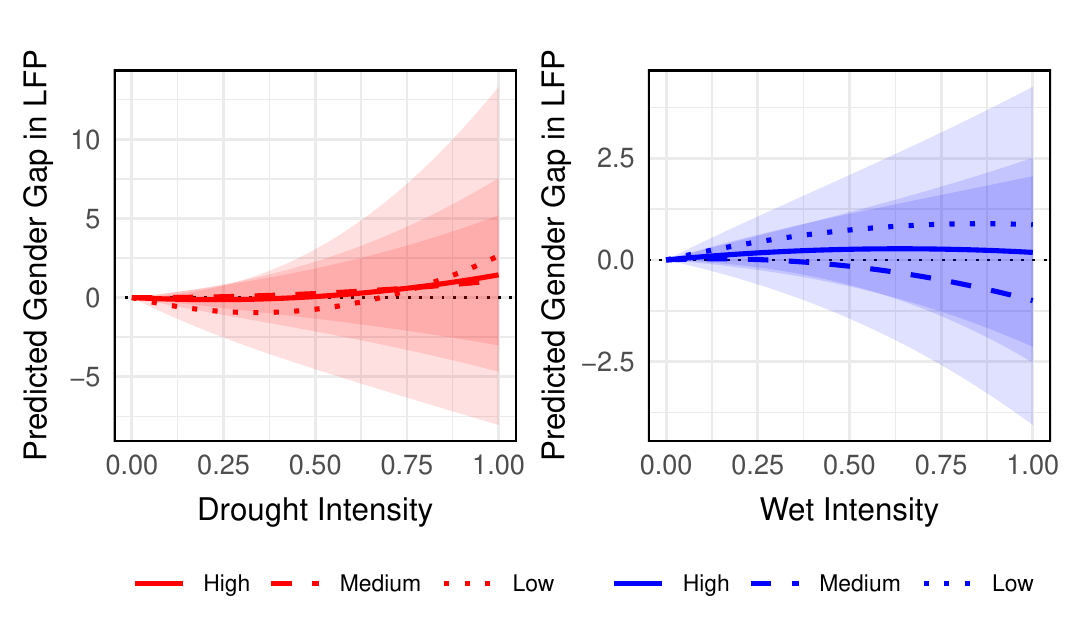}
    \caption{\textit{Heterogeneous Effects of Drought and Excess Wet Intensities on Gender Gap in LFP by Net Resilience. Predicted gender gap in LFP are plotted for different drought intensity (left-side panel) and excess wet intensity (right-side panel). Shaded areas indicate 95\% Confidence Intervals.}}
    \label{fig:het3}
\end{figure}

In countries with low resilience, extreme climate conditions matter more for gender gap in paid labour. Drought conditions share U-shaped relationship with gender gap in paid labour in low resilient countries (Column~(1), Table~\ref{tab:climate_gendergap_ndgain_gain} and Figure~\ref{fig:het3}). At low to moderate drought intensities, gender gap shrinks as women enter in paid labour to smooth household consumption and alleviate income loss \citep{kochar1999smoothing, Moshoeshoe2025}. But as drought intensity exceeds its turning point ($D^*=0.34$), gender gap expands in low resilient countries. Furthermore, wet intensity has linear but again positive influence on gender gap in LFP in countries with least resilience against climate change (Column~(1), Table~\ref{tab:climate_gendergap_ndgain_gain} and Figure~\ref{fig:het3}). 

The positive relationship between climate extremes and gender gap in countries with low resilience could be attributable to lack of readiness and limited adaptation capabilities of such countries. Agricultural sector tend to be predominant in less resilient economies where women are overrepresented \citep{fao2011state, sellers2024climate, palacios2017labor}. As droughts or excess wetness worsen, time required for unpaid domestic chores (e.g., collection of fuelwood and water) and care work increases \citep{sellers2024climate, unwomen2023climate}. These pull out women from participating in paid labour \citep{AfridiMahajanSangwan2022GenderedEffectsDroughts, 
fruttero2024gendered}. Moreover, structural and social norms also deter women to migrate away from affected regions and look for alternative income opportunities \citep{AfridiMahajanSangwan2022GenderedEffectsDroughts, sultana2024climate, sellers2024climate}.
\newpage

\section{Conclusion and Discussions}\label{conclusion}

This paper examines how extreme climate conditions shape gender disparity in labour force participation across 151 countries during past 24 years. The theoretical model shows that the balance between income and substitution effects determine whether adverse shock such as climate extremes expand or contract the gender gap in paid labour force participation. Key findings highlight non-linearities: first, gender gap in LFP exhibits U-shaped relationship with drought intensity. That is, gender gap narrows at low to moderate drought intensities but it starts to widen during extreme droughts. On the contrary, gender differences in paid labour shares an inverted U-shaped relationship with wet intensity. Thus, gender gap initially expands but declines once wet conditions become severe. Furthermore, drought-induced gender inequalities is mainly driven by gender gap in employment while wetness led gender gaps is attributable to gender gap in unemployment (ie., those seeking jobs). 

Country characteristics and institutional context play important role in shaping these relationships. Non-linear effects of excess wetness is most pronounced in countries with high disaster-related displacement risk. This is plausible because extreme wet shocks such as floods are sudden and cause widespread destruction which are likely to trigger displacement and influence gendered labour market outcomes. But droughts have non-linear impact on gender gap in countries with moderate disaster-related displacement risk where slow onset of droughts and gendered migration constraints gradually worsen gender gap in paid labour. Moreover, drought-driven non-linear effects are also concentrated in countries with least and moderate regulatory and legal support for women's empowerment. The inverted U-shaped relationship Limited agency and conservative institutional context often constrain women's mobility, asset ownership and participation in paid labour. Lastly, a country's resilience against climate shocks matters. The U-shaped nexus between gender gap in LFP and drought intensity is prominent in countries with least net resilience. Excess wetness also has positive effect on gender gap in paid labour in countries with weak resilience against climate extremes. 

These results have significant implications for climate adaptation policies. First, climate adaptation cannot assume uniform effects of climate extremes: same climate event such as droughts could narrow gender gaps at low intensity while expand the gender differences at high levels. Policies should adhere to the intensity of the shocks. Nature of climate event also matters for designing adaptation policies. Excess wetness and droughts have different types of relationship with gender inequality in paid labour. Second, strengthening institutional regulation towards women economic empowerment is important to reshape the extent to which climate extremes impact gender gap in paid labour. Countries with stronger women-friendly laws and regulations depict insignificant effect of droughts and wetness on gender gap in paid labour. Third, ecologically fragile countries should boost up their investments in infrastructure for climate actions: gender gap in LFP is not influenced by droughts and excess wetness in countries with moderate and high net resilience against climate events. To sum up, social insurance (e.g., unemployment benefits), social assistance (e.g., cash transfers) and care services (e.g., child care) should be provided to affected households in climate hotspots. This could help households to alleviate financial loss without forcing women to exit from labour market or enter distress-driven paid labour. 

Some limitations remain. Because this study utilises country-year panel data, I measure climatic moisture conditions by annual mean of SPEI-12 values. This approach could smooth within-year climate variations and weaken effects of short-lived extreme weather events in a country. Besides, the ILO modeled labour market series harmonise national estimates from labour force surveys of various countries. But the series may still have measurement error from differences in various national labour force surveys.

%\section*{Data availability statement}
%\WDDataStatement 

%\section*{Acknowledgements}
%During the preparation of this work the author used ChatGPT in order to crosscheck code, check for typos, and point out grammatical mistakes. After using this tool/service, the author reviewed and edited the content as needed and take full responsibility for the content of the published article.

%\section*{Declaration of interest}
%The author reports that there are no competing interests to declare.

\clearpage
\printbibliography

\clearpage
\appendix
\section*{Appendix}
\numberwithin{equation}{section}
\section{Theoretical Framework (Proof)}
\label{app:theory}

\paragraph{Setup.} 
Consider a representative household with husband ($m$) and wife ($f$). Each allocates one unit of time across paid work $L_j$, leisure $\ell_j$, and home production $h_j$:
\begin{equation}\label{eq:app_time}
L_j+\ell_j+h_j=1,\qquad j\in\{f,m\}.
\end{equation}

Preferences are
\begin{align}
U_m &= U_m(C_m,\ell_m,H), \label{eq:app_Um}\\
U_f &= U_f(C_f,\ell_f,H)-\kappa(L_f;\theta), \label{eq:app_Uf}
\end{align}
where $U_i$ is increasing and concave in its arguments. The term $\kappa(L_f;\theta)$ captures the social cost of women’s paid work, which varies with climate conditions $\theta$. A higher $\theta$ denotes a more adverse climate shock.

\paragraph{Social cost.}
I parameterize the social cost as
\begin{equation}\label{eq:app_kappa}
\kappa(L_f;\theta)=s(\theta)\,\tilde{\kappa}\,L_f,
\qquad s(\theta)>0,\quad s'(\theta)\ge 0,\quad \tilde{\kappa}>0,
\end{equation}
so the marginal social cost of women’s market work is
\begin{equation}\label{eq:app_kappaL}
\kappa_L(L_f;\theta)\equiv \frac{\partial \kappa(L_f;\theta)}{\partial L_f}=s(\theta)\,\tilde{\kappa}.
\end{equation}

\paragraph{Budget constraint.}
Household consumption is financed by labour income and non-labour income:
\begin{equation}\label{eq:app_budget0}
C_f+C_m=w_f L_f+w_m L_m+Y(\theta),
\end{equation}
where $Y'(\theta)<0$ and wages may also decline under adverse climate conditions, $w_j'(\theta)<0$.

\paragraph{Home production.}
The home good $H$ is produced using spouses' home time:
\begin{equation}\label{eq:app_H}
H=H(h_m,h_f;\theta), \qquad H_{h_j}>0,\;\; H_{h_jh_j}<0.
\end{equation}

\paragraph{Collective Bargaining Model.}
The household maximises a weighted sum of individual utilities:
\begin{equation}\label{eq:app_Uagg}
U=\alpha U_f+(1-\alpha)U_m,
\qquad \alpha\in(0,1),
\end{equation}
subject to \eqref{eq:app_time}, \eqref{eq:app_budget0}, and \eqref{eq:app_H}.

\paragraph{Rewriting the budget constraint}
Using $L_j=1-\ell_j-h_j$ from \eqref{eq:app_time}, rewrite \eqref{eq:app_budget0} as
\begin{equation}\label{eq:app_budget}
C_f+C_m+w_f(\ell_f+h_f)+w_m(\ell_m+h_m)=w_f+w_m+Y(\theta).
\end{equation}
\paragraph{Lagrangian Function}
Let $\lambda$ denote the multiplier on \eqref{eq:app_budget}. The Lagrangian is
\begin{align}
\mathcal{L}
&=
\alpha\Big[U_f(C_f,\ell_f,H)-\kappa(1-\ell_f-h_f;\theta)\Big]
+(1-\alpha)U_m(C_m,\ell_m,H)\notag\\
&\quad+\lambda\Big[w_f+w_m+Y(\theta)-C_f-C_m-w_f(\ell_f+h_f)-w_m(\ell_m+h_m)\Big].
\label{eq:app_lagrangian}
\end{align}

\paragraph{First-order conditions}
Denote by $u_H$ the \emph{marginal utility of household production} in the weighted objective:
\begin{equation}\label{eq:app_uH_def}
u_H \;\equiv\; \alpha\,U_{f,H}(C_f,\ell_f,H)+(1-\alpha)\,U_{m,H}(C_m,\ell_m,H).
\end{equation}

\textbf{Consumption.}
\begin{align}
\frac{\partial \mathcal{L}}{\partial C_f}=0
&\Rightarrow \alpha\,U_{f,C}=\lambda, \label{eq:app_foc_Cf}\\
\frac{\partial \mathcal{L}}{\partial C_m}=0
&\Rightarrow (1-\alpha)\,U_{m,C}=\lambda. \label{eq:app_foc_Cm}
\end{align}

\textbf{Home production time.}
Using $H_{h_j}\equiv \partial H/\partial h_j$ and $u_H=[\alpha\frac{\partial U_f}{\partial H}+(1-\alpha)\frac{\partial U_m}{\partial H}]$:
\begin{align}
\frac{\partial \mathcal{L}}{\partial h_f}=0
&\Rightarrow u_H\,H_{h_f}=\lambda w_f-\alpha\,\tilde{\kappa}, \label{eq:app_foc_hf}\\
\frac{\partial \mathcal{L}}{\partial h_m}=0
&\Rightarrow u_H\,H_{h_m}=\lambda w_m. \label{eq:app_foc_hm}
\end{align}

\paragraph{Opportunity costs of home production}
Define the shadow value of an hour in home production by spouse $j$:
\begin{equation}\label{eq:app_shadow_home}
q_j(\theta)\equiv \frac{u_H}{\lambda}\,H_{h_j}(h_m,h_f;\theta), \qquad j\in\{f,m\}.
\end{equation}

Using \eqref{eq:app_foc_hf}--\eqref{eq:app_foc_hm}, I obtain
\begin{align}
q_f(\theta)
&= w_f-\frac{\alpha}{\lambda}\,\kappa_L(L_f;\theta)
= w_f-\frac{\alpha}{\lambda}\,s(\theta)\tilde{\kappa},
\label{eq:app_qf}\\
q_m(\theta)
&= w_m.
\label{eq:app_qm}
\end{align}
Thus, women’s shadow value of home time equals their market wage net of the marginal social cost from working outside their home, whereas men’s shadow value of home time equals their wage.
The ratio of their opportunity cost is:
\begin{equation}\label{eq:app_home_ratio}
\frac{q_f(\theta)}{q_m(\theta)}
=
\frac{w_f-\frac{\alpha}{\lambda}s(\theta)\tilde{\kappa}}{w_m}.
\end{equation}

\paragraph{Reduced-form labour supply and income--substitution decomposition}
Solving the household problem yields Marshallian (uncompensated) labour supplies:
\begin{align}
L_f^* &= \mathcal{L}_f\!\big(Y(\theta), w_f(\theta), w_m(\theta), \tilde{\kappa} s(\theta), \theta, \alpha\big), \label{eq:app_Lfstar}\\
L_m^* &= \mathcal{L}_m\!\big(Y(\theta), w_f(\theta), w_m(\theta), \theta, \alpha\big). \label{eq:app_Lmstar}
\end{align}

Taking total derivatives with respect to $\theta$ gives the decomposition into an income and substitution effects:
\begin{align}
\frac{dL_f^*}{d\theta}
&=
\frac{\partial \mathcal{L}_f}{\partial Y}\,Y'(\theta)
+
\left[
\frac{\partial \mathcal{L}_f}{\partial w_f}\,w_f'(\theta)
+\frac{\partial \mathcal{L}_f}{\partial w_m}\,w_m'(\theta)
+\frac{\partial \mathcal{L}_f}{\partial s}\,s'(\theta)
+\frac{\partial \mathcal{L}_f}{\partial \theta}
\right],
\label{eq:app_income_sub_f}\\[6pt]
\frac{dL_m^*}{d\theta}
&=
\frac{\partial \mathcal{L}_m}{\partial Y}\,Y'(\theta)
+
\left[
\frac{\partial \mathcal{L}_m}{\partial w_f}\,w_f'(\theta)
+\frac{\partial \mathcal{L}_m}{\partial w_m}\,w_m'(\theta)
+\frac{\partial \mathcal{L}_m}{\partial \theta}
\right].
\label{eq:app_income_sub_m}
\end{align}

\textbf{Gender gap response.} Effect of the shock on gender gap in paid labour is given by $\mathrm{Gap}\equiv L_m^*-L_f^*$, then
,
\begin{equation}\label{eq:app_gap}
\begin{aligned}
\frac{d\,\mathrm{Gap}}{d\theta} &= \frac{dL_m^*}{d\theta} - \frac{dL_f^*}{d\theta} \\[10pt]
\implies \frac{d\,\mathrm{Gap}}{d\theta} &= \left[ \frac{\partial \mathcal{L}_m}{\partial Y} - \frac{\partial \mathcal{L}_f}{\partial Y} \right] Y'(\theta) \\
&\quad + \left[ \frac{\partial (\mathcal{L}_m - \mathcal{L}_f)}{\partial w_f} w_f'(\theta) + \frac{\partial (\mathcal{L}_m - \mathcal{L}_f)}{\partial w_m} w_m'(\theta) \right] \\
&\quad - \frac{\partial \mathcal{L}_f}{\partial s} s'(\theta) + \left[ \frac{\partial \mathcal{L}_m}{\partial \theta} - \frac{\partial \mathcal{L}_f}{\partial \theta} \right]\\
\implies \frac{d\,\mathrm{Gap}}{d\theta} &=AY'(\theta)+Bw'_f(\theta)+Cw'_m(\theta)- \frac{\partial \mathcal{L}_f}{\partial s} s'(\theta) +D(\theta)
\end{aligned}
\end{equation}

\noindent where $Y'(\theta)<0, w'_f(\theta)<0, w'_m(\theta)<0, s'(\theta)>0$, $A=[ \frac{\partial \mathcal{L}_m}{\partial Y} - \frac{\partial \mathcal{L}_f}{\partial Y}]$, $B=[\frac{\partial (\mathcal{L}_m - \mathcal{L}_f)}{\partial w_f}$), $C=\frac{\partial (\mathcal{L}_m - \mathcal{L}_f)}{\partial w_m}$ and $D=\left[ \frac{\partial \mathcal{L}_m}{\partial \theta} - \frac{\partial \mathcal{L}_f}{\partial \theta} \right]$.

\newpage
\section{Main Regression Results}
\begin{table}[!htbp]\centering
\caption{Climate Extremes and Gender Gap in Paid Labour Outcomes}
\label{tab:climate_gendergaps_checks}
\begin{threeparttable}
\small
\begin{tabular}{lccc}
\toprule
Gender gap & LFP & Employment & Unemployment \\
 & (1) & (2) & (3) \\
\midrule
$\mathrm{Drought}^2_{ct}$   & 5.354$^{***}$ & 4.088$^{***}$ & 1.037 \\
                            & (1.431)       & (1.417)       & (1.096) \\
$\mathrm{Drought}_{ct}$     & -2.977$^{**}$ & -2.447$^{**}$ & -0.230 \\
                            & (1.187)       & (1.187)       & (0.768) \\
$\mathrm{Wet}_{ct}^2$       & -1.119$^{*}$  & -0.617        & -0.938$^{**}$ \\
                            & (0.628)       & (0.653)       & (0.465) \\
$\mathrm{Wet}_{ct}$         & 1.369$^{*}$   & 0.903         & 0.628 \\
                            & (0.718)       & (0.720)       & (0.398) \\
\addlinespace
\midrule
Economic growth (lag)                 & $\checkmark$ & $\checkmark$ & $\checkmark$ \\
Urbanization (lag)                    & $\checkmark$ & $\checkmark$ & $\checkmark$ \\
FDI inflows (lag, \% of GDP)          & $\checkmark$ & $\checkmark$ & $\checkmark$ \\
Age dependency                        & $\checkmark$ & $\checkmark$ & $\checkmark$ \\
Fertility                             & $\checkmark$ & $\checkmark$ & $\checkmark$ \\
Sex ratio                             & $\checkmark$ & $\checkmark$ & $\checkmark$ \\
Population density                    & $\checkmark$ & $\checkmark$ & $\checkmark$ \\
\addlinespace
Country FE                            & Yes & Yes & Yes \\
Year FE                               & Yes & Yes & Yes \\
\midrule
Observations                          & 3,646 & 3,646 & 3,646 \\
Countries (clusters)                  & 156   & 156   & 156   \\
$R^{2}$                               & 0.967 & 0.964 & 0.789 \\
Within $R^{2}$                        & 0.045 & 0.039 & 0.036 \\
\bottomrule
\end{tabular}
\begin{tablenotes}\footnotesize
\item Notes: Dependent variables are gender gaps in labour force participation, employment rate, and unemployment rate. $\mathrm{Drought}_{ct}$ and $\mathrm{Wet}_{ct}$ are climate extreme measures constructed from 12-month SPEI tail-ends on both sides (see text); squared terms allow for non-linear effects. Robust standard errors clustered at the country level are in parentheses. $^{***}p<0.01$, $^{**}p<0.05$, $^{*}p<0.10$.
\end{tablenotes}
\end{threeparttable}
\end{table}
\newpage
\section{Regression Results for Robustness Checks}
\begin{table}[!htbp]\centering
\caption{Robustness: Spline Specification}
\label{tab:robust_spline_3cols}
\begin{threeparttable}
\small
\begin{tabular}{lccc}
\toprule
  & {LFP} & {Employment} & {Unemployment} \\
 & (1) & (2) & (3) \\
\midrule
$\mathrm{DroughtSpline}_{ct,1}$ (dcs1)     & -3.361$^{**}$ & -2.898$^{**}$ & 0.063 \\
                                          & (1.306)       & (1.259)       & (0.871) \\
$\mathrm{DroughtSpline}_{ct,2}$ (dcs2)     &  5.501$^{***}$ &  4.426$^{***}$ & 0.534 \\
                                          & (1.577)        & (1.520)        & (1.212) \\
$\mathrm{WetSpline}_{ct,1}$ (wcs1)         &  1.957$^{**}$ &  1.143         & 1.098$^{*}$ \\
                                          & (0.870)       & (0.913)        & (0.577) \\
$\mathrm{WetSpline}_{ct,2}$ (wcs2)         & -2.154$^{**}$ & -1.075         & -1.764$^{**}$ \\
                                          & (0.976)       & (1.048)        & (0.781) \\
\midrule
Economic growth (lag)                      & \checkmark & \checkmark & \checkmark \\
Urbanization (lag)                         & \checkmark & \checkmark & \checkmark \\
FDI inflows (lag, \% of GDP)               & \checkmark & \checkmark & \checkmark \\
Age dependency                             & \checkmark & \checkmark & \checkmark \\
Fertility                                  & \checkmark & \checkmark & \checkmark \\
Sex ratio                                  & \checkmark & \checkmark & \checkmark \\
Population density                         & \checkmark & \checkmark & \checkmark \\
\midrule
Country FE                                 & Yes & Yes & Yes \\
Year FE                                    & Yes & Yes & Yes \\
\midrule
Observations                               & 3,646 & 3,646 & 3,646 \\
Countries (clusters)                       & 156 & 156 & 156 \\
$R^2$                                      & 0.967 & 0.964 & 0.789 \\
Within $R^2$                               & 0.046 & 0.040 & 0.036 \\
\bottomrule
\end{tabular}
\begin{tablenotes}\footnotesize
\item Notes: Dependent variables are gender gaps (male minus female, in percentage points) in labour force participation (LFP), employment rate, and unemployment rate. Drought and wetness tail-intensity measures derived from SPEI-12 are modeled flexibly using restricted cubic spline bases. Knots are set at the 25th, 50th, and 75th percentiles of the positive tail intensity distributions: drought knots $(0.106,\,0.239,\,0.416)$ and wet knots $(0.114,\,0.271,\,0.488)$. All specifications include country and year fixed effects and the full set of controls shown. Robust standard errors clustered at the country level are in parentheses. $^{***}p<0.01$, $^{**}p<0.05$, $^{*}p<0.10$.
\end{tablenotes}
\end{threeparttable}
\end{table}
\newpage
\begin{table}[!htbp]\centering
\caption{Robustness: Piecewise Linear Specification}
\label{tab:robust_piecewise_3cols}
\begin{threeparttable}
\small
\begin{tabular}{lccc}
\toprule
 & {LFP} & {Employment} & {Unemployment} \\
 & (1) & (2) & (3) \\
\midrule
$\mathrm{DroughtBelow}_{ct}$ ($\min\{\mathrm{Drought}_{ct},D^{*}\}$) & -2.800$^{**}$ & -2.380$^{**}$ & 0.056 \\
                                                                    & (1.147)       & (1.119)       & (0.745) \\
$\mathrm{DroughtAbove}_{ct}$ ($\max\{0,\mathrm{Drought}_{ct}-D^{*}\}$) &  3.732$^{***}$ &  2.748$^{***}$ & 0.801 \\
                                                                      & (0.967)        & (0.974)        & (0.786) \\
$\mathrm{WetBelow}_{ct}$ ($\min\{\mathrm{Wet}_{ct},W^{*}\}$)         &  0.799         &  0.567         & 0.233 \\
                                                                    & (0.498)       & (0.479)       & (0.228) \\
$\mathrm{WetAbove}_{ct}$ ($\max\{0,\mathrm{Wet}_{ct}-W^{*}\}$)       & -0.857         & -0.263         & -1.464$^{**}$ \\
                                                                    & (0.796)       & (0.829)       & (0.682) \\
\midrule
Economic growth (lag)                      & \checkmark & \checkmark & \checkmark \\
Urbanization (lag)                         & \checkmark & \checkmark & \checkmark \\
FDI inflows (lag, \% of GDP)               & \checkmark & \checkmark & \checkmark \\
Age dependency                             & \checkmark & \checkmark & \checkmark \\
Fertility                                  & \checkmark & \checkmark & \checkmark \\
Sex ratio                                  & \checkmark & \checkmark & \checkmark \\
Population density                         & \checkmark & \checkmark & \checkmark \\
\midrule
Country FE                                 & Yes & Yes & Yes \\
Year FE                                    & Yes & Yes & Yes \\
\midrule
Observations                               & 3,646 & 3,646 & 3,646 \\
Countries (clusters)                       & 156 & 156 & 156 \\
$R^2$                                      & 0.967 & 0.964 & 0.789 \\
Within $R^2$                               & 0.045 & 0.039 & 0.036 \\
\bottomrule
\end{tabular}
\begin{tablenotes}\footnotesize
\item Notes: Dependent variables are gender gaps (male minus female, in percentage points) in labour force participation (LFP), employment rate, and unemployment rate. $\mathrm{Drought}_{ct}$ and $\mathrm{Wet}_{ct}$ are tail-intensity measures derived from SPEI-12 using the 10th and 90th percentile thresholds. The piecewise linear specification allows different slopes below and above knots $D^{*}=0.278$ and $W^{*}=0.612$. All specifications include country and year fixed effects and the full set of controls shown. Robust standard errors clustered at the country level are in parentheses. $^{***}p<0.01$, $^{**}p<0.05$, $^{*}p<0.10$.
\end{tablenotes}
\end{threeparttable}
\end{table}

\newpage
\section{Regression Results for Heterogeneous Effects}
%Disaster Displacement Risk
% ============================================================
% Heterogeneous effects by displacement (Overleaf-ready)
% disp_het: 1=Low, 2=Medium, 3=High (terciles of displacement_figure)
% ============================================================
\begin{table}[!htbp]\centering
\caption{Climate Extremes and Gender Gap in Labour Force Participation by Natural Disaster related Displacement Risk}
\label{tab:climate_gendergap_disphet}
\begin{threeparttable}
\small
\begin{tabular}{lccc}
\toprule
Displacement Risk & Low & Medium & High \\
 & (1) & (2) & (3) \\
\midrule
$\mathrm{Drought}^2_{ct}$        & -0.446        & 6.952$^{**}$  & 9.862 \\
                                & (2.352)       & (3.107)       & (6.099) \\

$\mathrm{Drought}_{ct}$          & -0.156        & -4.772        & -5.903 \\
                                & (1.814)       & (2.911)       & (3.793) \\
$\mathrm{Wet}^2_{ct}$            & -1.171        & -0.856        & -9.500$^{**}$ \\
                                & (0.827)       & (1.321)       & (4.166) \\
$\mathrm{Wet}_{ct}$              & 1.761         & 0.447         & 7.405$^{***}$ \\
                                & (1.232)       & (1.424)       & (2.732) \\
\addlinespace
\midrule
Economic growth (lag)            & $\checkmark$  & $\checkmark$  & $\checkmark$ \\
Urbanization (lag)               & $\checkmark$  & $\checkmark$  & $\checkmark$ \\
FDI inflows (lag, \% of GDP)     & $\checkmark$  & $\checkmark$  & $\checkmark$ \\
Age dependency                   & $\checkmark$  & $\checkmark$  & $\checkmark$ \\
Fertility                        & $\checkmark$  & $\checkmark$  & $\checkmark$ \\
Sex ratio                        & $\checkmark$  & $\checkmark$  & $\checkmark$ \\
Population density               & $\checkmark$  & $\checkmark$  & $\checkmark$ \\
\addlinespace
Country FE                       & Yes           & Yes           & Yes \\
Year FE                          & Yes           & Yes           & Yes \\
\midrule
Observations                     & 999           & 1,008         & 1,111 \\
Countries (clusters)             & 43            & 42            & 48 \\
$R^{2}$                          & 0.948         & 0.957         & 0.981 \\
Within $R^{2}$                   & 0.068         & 0.125         & 0.121 \\
\bottomrule
\end{tabular}
\begin{tablenotes}\footnotesize
\item \textit{Notes: Dependent variable is the gender gap in labour force participation (percentage points). Using the natural disaster displacement risks, countries are classified into terciles based on the sample distribution:
low risk (\texttt{displacement}$\le$ 1{,}131), medium risk (1{,}407 $\le$ \texttt{displacement} $\le$ 29{,}542), and high risk (\texttt{displacement} $\ge$ 31{,}164). Robust standard errors clustered at the country level are in parentheses. $^{***}p<0.01$, $^{**}p<0.05$, $^{*}p<0.10$.}
\end{tablenotes}
\end{threeparttable}
\end{table}

%Women Empowerment
% ============================================================
% Heterogeneous effects by women's empowerment (Overleaf-ready)
% women_het: 1=Low, 2=Medium, 3=High (3 quantiles of WBL index)
% ============================================================
\begin{table}[!htbp]\centering
\caption{Climate Extremes and Gender Gap in Labour Force Participation by Women's Empowerment}
\label{tab:climate_gendergap_empowerhet}
\begin{threeparttable}
\small
\begin{tabular}{lccc}
\toprule
Empowerment group & Low & Medium & High \\
 & (1) & (2) & (3) \\
\midrule
$\mathrm{Drought}^2_{ct}$        & 8.634$^{**}$  & 1.698$^{**}$  & 2.997 \\
                                & (3.454)       & (0.742)       & (1.965) \\
$\mathrm{Drought}_{ct}$          & -4.914$^{**}$ & -0.667        & -1.763 \\
                                & (2.378)       & (0.747)       & (1.486) \\
$\mathrm{Wet}^2_{ct}$            & -0.407        & -1.272$^{**}$ & -1.167 \\
                                & (0.760)       & (0.596)       & (1.316) \\
$\mathrm{Wet}_{ct}$              & 1.526         & 1.005         & 1.202 \\
                                & (1.119)       & (0.695)       & (1.124) \\
\addlinespace
\midrule
Economic growth (lag)            & $\checkmark$  & $\checkmark$  & $\checkmark$ \\
Urbanization (lag)               & $\checkmark$  & $\checkmark$  & $\checkmark$ \\
FDI inflows (lag, \% of GDP)     & $\checkmark$  & $\checkmark$  & $\checkmark$ \\
Age dependency                   & $\checkmark$  & $\checkmark$  & $\checkmark$ \\
Fertility                        & $\checkmark$  & $\checkmark$  & $\checkmark$ \\
Sex ratio                        & $\checkmark$  & $\checkmark$  & $\checkmark$ \\
Population density               & $\checkmark$  & $\checkmark$  & $\checkmark$ \\
\addlinespace
Country FE                       & Yes           & Yes           & Yes \\
Year FE                          & Yes           & Yes           & Yes \\
\midrule
Observations                     & 1,141         & 1,187         & 1,254 \\
Countries (clusters)             & 73            & 99            & 84 \\
$R^{2}$                          & 0.985         & 0.982         & 0.965 \\
Within $R^{2}$                   & 0.192         & 0.054         & 0.191 \\
\bottomrule
\end{tabular}
\begin{tablenotes}\footnotesize
\item \textit{Notes: Dependent variable is the gender gap in labour force participation (percentage points). Using the Women, Business and the Law (WBL) index, countries are classified into terciles of women’s empowerment based on the sample distribution: low empowerment (WBL $\le$ 58.75), medium empowerment (59.375 $\le$ WBL $\le$ 75.0), and high empowerment (WBL $\ge$ 75.625).
Robust standard errors clustered at the country level are in parentheses. $^{***}p<0.01$, $^{**}p<0.05$, $^{*}p<0.10$.}
\end{tablenotes}
\end{threeparttable}
\end{table}

% ============================================================
% Heterogeneous effects by Net Resilience
% ============================================================
%Net Gain
% ============================================================
% Heterogeneous effects by ND-GAIN (GAIN) terciles (Overleaf-ready)
% gain_tertile: 1=Low, 2=Medium, 3=High (terciles of ND-GAIN GAIN)
% ============================================================
\begin{table}[!htbp]\centering
\caption{Climate Extremes and Gender Gap in Labour Force Participation by Net Resilience}
\label{tab:climate_gendergap_ndgain_gain}
\begin{threeparttable}
\small
\begin{tabular}{lccc}
\toprule
Net Resilience & Low & Medium & High \\
 & (1) & (2) & (3) \\
\midrule
$\mathrm{Drought}^2_{ct}$        & 8.215$^{*}$   & 1.156         & 2.612 \\
                                & (4.354)       & (1.514)       & (2.450) \\
$\mathrm{Drought}_{ct}$          & -5.580$^{*}$  & -0.071        & -1.178 \\
                                & (3.105)       & (1.396)       & (1.835) \\
$\mathrm{Wet}^2_{ct}$            & -1.207        & -1.376        & -0.676 \\
                                & (1.181)       & (0.956)       & (0.831) \\
$\mathrm{Wet}_{ct}$              & 2.073$^{*}$   & 0.378         & 0.861 \\
                                & (1.215)       & (1.198)       & (0.811) \\
\addlinespace
\midrule
Economic growth (lag)            & $\checkmark$  & $\checkmark$  & $\checkmark$ \\
Urbanization (lag)               & $\checkmark$  & $\checkmark$  & $\checkmark$ \\
FDI inflows (lag, \% of GDP)     & $\checkmark$  & $\checkmark$  & $\checkmark$ \\
Age dependency                   & $\checkmark$  & $\checkmark$  & $\checkmark$ \\
Fertility                        & $\checkmark$  & $\checkmark$  & $\checkmark$ \\
Sex ratio                        & $\checkmark$  & $\checkmark$  & $\checkmark$ \\
Population density               & $\checkmark$  & $\checkmark$  & $\checkmark$ \\
\addlinespace
Country FE                       & Yes           & Yes           & Yes \\
Year FE                          & Yes           & Yes           & Yes \\
\midrule
Observations                     & 1,174         & 1,110         & 1,179 \\
Countries (clusters)             & 64            & 73            & 64 \\
$R^{2}$                          & 0.977         & 0.976         & 0.972 \\
Within $R^{2}$                   & 0.151         & 0.042         & 0.190 \\
\bottomrule
\end{tabular}
\begin{tablenotes}\footnotesize
\item Notes: Dependent variable is the gender gap in labour force participation (percentage points). Using the ND-GAIN's \textit{GAIN} index, countries are classified into terciles of net resilience based on the sample distribution:
low resilience (GAIN $\le$ 39.741), middle resilience (39.746 $\le$ GAIN $\le$ 49.026), and high resilience (GAIN $\ge$ 49.033). Robust standard errors clustered at the country level are in parentheses. $^{***}p<0.01$, $^{**}p<0.05$, $^{*}p<0.10$.
\end{tablenotes}
\end{threeparttable}
\end{table}

\end{document}